\documentclass{CUP-JNL-DTM}%

\usepackage{graphicx}
\usepackage{multicol,multirow}
\usepackage{amsmath,amssymb,amsfonts}
\usepackage{mathrsfs}
\usepackage{amsthm}
\usepackage{rotating}
\usepackage{appendix}
\usepackage[numbers]{natbib}
\usepackage{ifpdf}
\usepackage[T1]{fontenc}
\usepackage{newtxtext}
\usepackage{newtxmath}
\usepackage{textcomp}
\usepackage{xcolor}
\usepackage{lipsum}
\usepackage[colorlinks,allcolors=blue]{hyperref}
\usepackage{caption}

\newcommand{\z}{\textbf{z}}
\newcommand{\x}{\textbf{x}}

\theoremstyle{definition}

\numberwithin{equation}{section}

\jname{Environmental Data Science}
\articletype{Application Paper}
\jyear{2023}

\begin{document}

\begin{Frontmatter}

\title[Understanding Precipitation Changes through Unsupervised Machine Learning]{Understanding Precipitation Changes through Unsupervised Machine Learning}

\author[1]{Griffin Mooers}
\author[2,3]{Tom Beucler}
\author[4,5]{Mike Pritchard}
\author[6]{Stephan Mandt}

\authormark{Mooers \textit{et al}.}

\address[1]{\orgdiv{Earth System Science}, \orgname{University of California}, \orgaddress{\city{Irvine}, \postcode{92617}, \state{CA},  \country{USA}}
\email{gmooers96@gmail.com}}

\address[2]{\orgdiv{Faculty of Geosciences and Environment}, \orgname{University of Lausanne}, \orgaddress{\city{Lausanne}, \postcode{1015}, \state{VD},  \country{Switzerland}} 
}

\address[3]{\orgdiv{Expertise Center for Climate Extremes}, \orgname{University of Lausanne}, \orgaddress{\city{Lausanne}, \postcode{1015}, \state{VD},  \country{Switzerland}} 
}

\address[4]{\orgdiv{Earth System Science}, \orgname{University of California}, \orgaddress{\city{Irvine}, \postcode{92617}, \state{CA},  \country{USA}}
}

\address[5]{\orgdiv{NVIDIA Research}, \orgaddress{\city{Santa Clara}, \postcode{95050}, \state{CA},  \country{USA}}}

\address[6]{\orgdiv{Department of Computer Science}, \orgname{University of California}, \orgaddress{\city{Irvine}, \postcode{92617}, \state{CA},  \country{USA}}}

\authormark{Mooers et al.}


\abstract{Despite the importance of quantifying how the spatial patterns of heavy precipitation will change with warming, we lack tools to objectively analyze the storm-scale outputs of modern climate models. To address this gap, we develop an unsupervised, spatial machine-learning framework to quantify how storm dynamics affect changes in heavy precipitation. 
We find that changes in heavy precipitation (above the 80th percentile) are predominantly explained by changes in the frequency of these events, rather than by changes in how these storm regimes produce precipitation. Our study shows how unsupervised machine learning, paired with domain knowledge, may allow us to better understand the physics of the atmosphere and anticipate the changes associated with a warming world.}

\end{Frontmatter}

\section*{Impact Statement}
Heavy precipitation causes severe flooding and water storage issues across the globe. A better understanding of how the heavy precipitation will change with warming is a priority. However, the details of the heavy precipitation are controlled by storm-scale atmospheric dynamics which are normally neglected or averaged out. We present a data-driven approach using machine learning to leverage the spatial details of dynamics controlling heavy precipitation. This allows us to determine the degree to which changes in convection type are responsible for shifts in heavy precipitation as the climate warms. 
With most machine learning focusing on prediction problems, our study highlights the impact of unsupervised machine learning for knowledge discovery in climate science.

\section{Introduction: Understanding the Changing Spatial Patterns of Heavy Precipitation}

According to the latest Intergovernmental Panel on Climate Change  report~\cite{douville2021water}, ``there is high confidence that heavier precipitation events across the globe will increase in both intensity and frequency with global warming''. As the severity of storms and tropical cyclones magnifies, there will be associated increases in flood-related risk~\cite{hess-22-2041-2018} and challenges in water management~\cite{Adger_2007, Adger_2009}. To first order, bounds of heavy precipitation are limited by the water vapor holding capacity of the atmosphere, which increases by about 7\% per 1K (Kelvin) of warming following an approximate Clausius-Clapeyron scaling~\cite{o2009scaling}. This is referred to as the ``thermodynamic contribution'' to heavy precipitation changes~\cite{emori2005dynamic} and gives a solid theoretical foundation for~\textit{spatially-averaged} changes in heavy precipitation. 

Yet climate change adaptation requires knowledge of how heavy precipitation will change at the~\textit{local} scale, i.e., understanding the~\textit{changing spatial patterns} of heavy precipitation under warming. Focusing on the tropics, where most of the vulnerable world population lives~\cite{edelman2014state}, these changing spatial patterns are primarily dictated by atmospheric vertical velocity (``dynamical'') changes because horizontal spatial gradients in temperatures are weak. This is referred to as the ``dynamic contribution'' to heavy precipitation changes~\cite{emori2005dynamic}. 

\begin{figure}[ht!]
\centering
\includegraphics[width=0.75\linewidth]{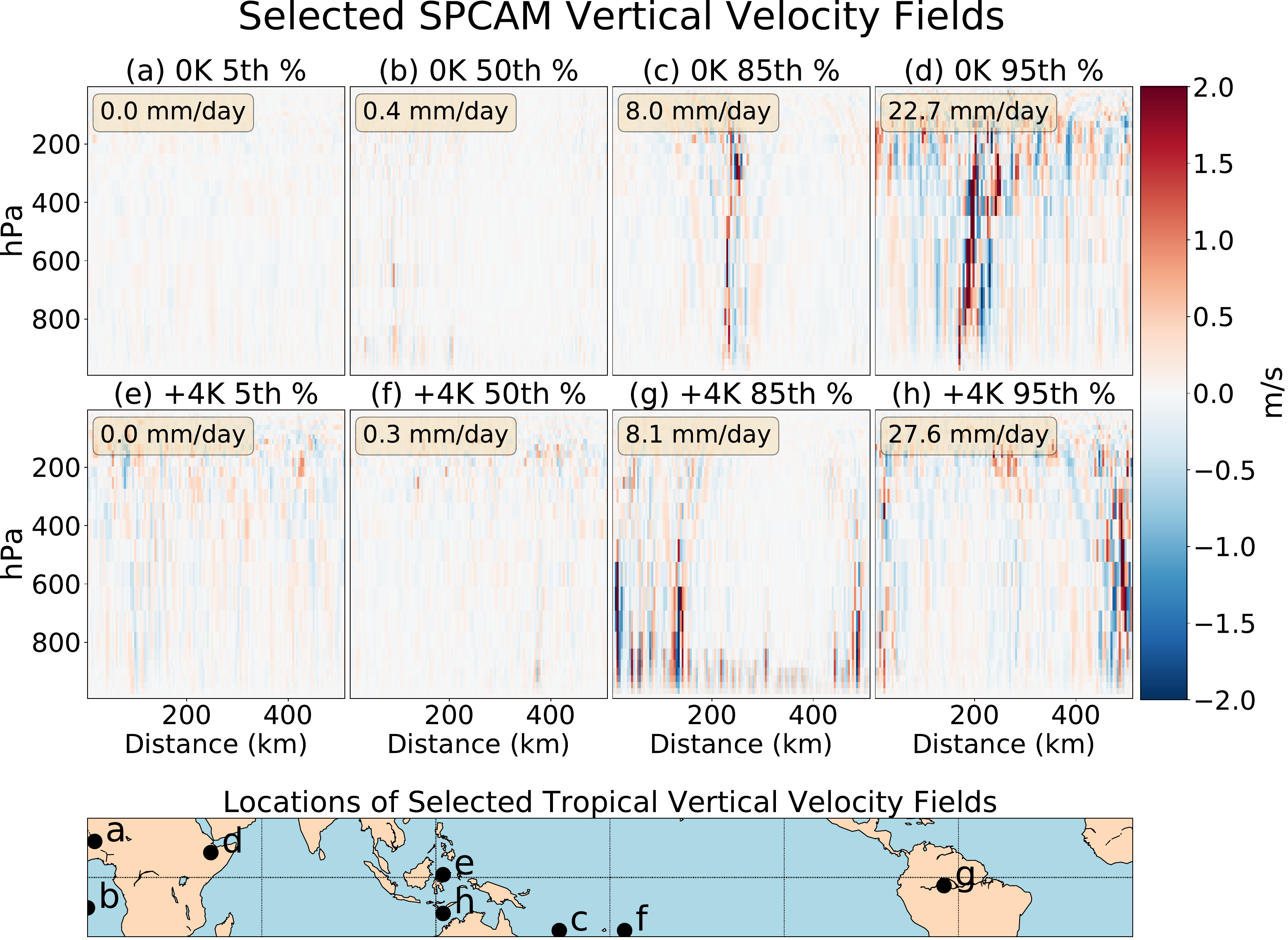}
\caption{\label{fig:W_Fields}Selected vertical velocity fields from our ``Control'' (0K, a-d) and ``Warmed'' (+4K, e-h) SPCAM simulations. By sampling the precipitation distribution, we show instances of vertical velocity fields associated with no precipitation (a, e), drizzle (b, f), heavy rainfall (c, g), and intense storms (d, h).}
\end{figure}

A comprehensive understanding of this ``dynamic contribution'' remains elusive. Approximate scalings can be derived based on quasi-geostrophic dynamics~\cite{li2020response,o2015precipitation} and convective storm dynamics~\cite{muller2011intensification,abbott2020convective}. As well, some of the effects from dynamics can be approximated using observational and reanalysis data~\cite{Tan_Nature}. But actionable findings require Earth-like simulations of the present and future climates (e.g.,~\cite{pendergrass2014changes}), which can resolve regional circulation changes and their effects on storms in their full complexity. These simulations are computationally demanding and output large amounts of multi-scale, three-dimensional data that challenge traditional data analysis tools. For example, the state-of-the-art storm-resolving\footnote{5 kilometers or less horizontal grid spacing} SPCAM (Super Parameterized Community Atmospheric Model~\cite{10.1175/1520-0469(2003)060<0607:CRMOTA>2.0.CO;2, Khairoutdinov1999}) simulations we use in this study (Section~\ref{sec:preprocessing}) output 3.4 Terabytes of data over 90 days, with 76,944,384 samples of precipitation and the corresponding storm-scale vertical velocity fields (see Fig~\ref{fig:W_Fields} for examples). 

\section{Theory: Heavy Precipitation Decomposition\label{sec:theory}}

In this section, we outline our strategy to facilitate the analysis of heavy precipitation changes including spatial details elucidating storm formation. 

The crux of this analysis rests on the assumption that we can meaningfully cluster different vertical velocity fields into $N$ different convection types. Our analysis transforms these convection types into “cluster probabilities”. For the total $N$ data points being assigned to a set of K clusters, we first compute a vector of counts $N_k$, indicating the number of data points assigned to each cluster index by $k$: $\sum_{k=1}^N N_k \equiv N$. We now convert these numbers to a normalized probability vector as $\pi \equiv (\pi_1, …, \pi_k) \equiv (N_1/N, …, N_k/N)$.

To calculate these $\pi_i$'s, we use variational autoencoders in conjunction with k-means clustering, also called vector quantization~\cite{gray1984vector,Lloyd_1957}. Details on the exact coarse graining procedure will be presented in Section~\ref{sec:training}; for now, we take the $N$ different convection types and their probabilities as given.

This unsupervised quantization of regimes of convection through ML allows us to quantitatively understand changes in heavy precipitation ($P_{\mathrm{heavy}}$) from both changes in convection regime characteristics and probability. Here we define $P_{\mathrm{heavy}} $ as a fixed~\textit{high} quantile of precipitation (e.g., 80th-99.99th percentiles) at a given spatial location. To model the effects of climate change, we define $\Delta P_{\mathrm{heavy}} $ as its absolute change from the ``Control'' to the ``Warmed'' climate, and show below that relative changes in heavy precipitation can be decomposed using changes in $\pi $ (Eq.~\ref{eq:dynamical_contribution}).

We derive a decomposition of heavy precipitation changes for global warming by making a series of simple physical assumptions about precipitation. While these approximations facilitate our results' interpretability and lend physical meaning to each term, they need not hold to derive Eq.~\ref{eq:P_extreme}, the final form of the decomposition as it could also be derived by decomposing the heavy precipitation field into its spatial-average and an anomaly, before further decomposing the anomaly using the objectively-identified dynamical regimes. To first order, precipitation ($P $) scales like condensation rate ($C $), which depends on the full vertical velocity ($w $) and atmospheric water vapor (here quantified using specific humidity $q $) fields:

\begin{equation}
P\approx C\left(w,q\right).
\label{eq:P_w_q}
\end{equation}

Note that Eq.~\ref{eq:P_w_q} neglects the dependence on microphysical processes (see e.g., \cite{muller2020response}) to focus on the thermodynamical and dynamical components of precipitation. When focusing on heavier precipitation, we de facto sample atmospheric columns that are so humid that the specific humidity $q $ equals its saturation value $q_{\mathrm{sat}}$. This allows us to further simplify Eq.~\ref{eq:P_w_q} in the case of high quantiles of $P $:

\begin{equation}
P_{\mathrm{heavy}}\approx P_{\mathrm{heavy}}\left(w,q_{\mathrm{sat}}\right).
\label{eq:Pextreme_w_qsat}
\end{equation}

We now make the assumption that the thermodynamic dependence on $q_{\mathrm{sat}} $ can be factored out of the right-hand side of Eq.~\ref{eq:Pextreme_w_qsat} and denote the dynamical pre-factor as ${\cal D}\left(w\right) $: 

\begin{equation}
P_{\mathrm{heavy}}\approx q_{\mathrm{sat}}\times {\cal D}\left(w\right).
\label{eq:isolate_dynamics}
\end{equation}

The previous assumption has been historically justified by assuming a moist adiabatic temperature profile and scaling the vertical velocity by using a single value for the entire vertical profile for intense events~\cite{o2009scaling,muller2011intensification}. However, it is more accurate to assume vertical velocity profiles collapse when changing the vertical coordinate from pressure to the normalized integral of the moisture lapse rate~\cite{abbott2020convective}. 

We can now linearly decompose the dynamical pre-factor ${\cal D}\left(w\right) $ into the $N $ regimes identified by our unsupervised learning framework: 

\begin{equation}
{\cal D}\left(w, P\right)\approx{\cal D}_{0}+\sum_{i=1}^{N}{\cal D}_{i}\cdot\pi_{i},
\label{eq:D}
\end{equation}
where $\pi_{i} $ is the normalized probability of each dynamical regime determined by the vertical velocity field information. Combining Eq.'s~\ref{eq:isolate_dynamics}, \ref{eq:D}, and taking a logarithmic derivative with respect to climate change allows us to decompose relative changes in upper precipitation quantiles as follows:  
\begin{equation}
\frac{\Delta P_{\mathrm{heavy}}}{P_{\mathrm{heavy}}}\approx\frac{\Delta q_{\mathrm{sat}}}{q_{\mathrm{sat}}}+\frac{\Delta\left({\cal D}_{0}+\sum_{i=1}^{N}\pi_{i}{\cal D}_{i}\right)}{{\cal D}_{0}+\sum_{i=1}^{N}\pi_{i}{\cal D}_{i}},
\label{eq:P_extreme}
\end{equation}
where $\Delta $ denotes absolute changes from the reference to the warm climate. Lastly, we approximate the thermodynamic contribution to upper quantile precipitation as the relative changes in near-surface saturation specific humidity, which can be further approximated as spatially uniform:
\begin{equation}
q_{\mathrm{sat}}=q_{\mathrm{sat}}\left(T_{s},p_{s}\right)\Rightarrow\frac{\Delta q_{\mathrm{sat}}}{q_{\mathrm{sat}}}\approx7\%,
\label{eq:q_sat}
\end{equation}
where $T_{s} $ is near-surface temperature and $p_{s} $ near-surface pressure. Expanding Eq.~\ref{eq:P_extreme} and substituting ${\cal D}\left(w\right) $ using Eq.~\ref{eq:isolate_dynamics} yields the desired decomposition of heavy precipitation changes in climate:

\begin{equation}
    \frac{\Delta P_{\mathrm{heavy}}}{P_{\mathrm{heavy}}}=\overbrace{\underbrace{\frac{\Delta q_{\mathrm{sat}}}{q_{\mathrm{sat}}}}_{\mathrm{From\ theory}}}^{\mathrm{Thermodynamic}}+\overbrace{\underbrace{\frac{q_{\mathrm{sat}}}{P_{\mathrm{heavy}}}}_{\mathrm{From\ current\ climate}}\left(\Delta{\cal D}_{0}+\underbrace{\sum_{i=1}^{N}\Delta\pi_{i}{\cal D}_{i}}_{\mathrm{Regime\ prob.\ shifts}}+\underbrace{\sum_{i=1}^{N}\pi_{i}\Delta{\cal D}_{i}}_{\mathrm{Intra-regime\ changes}}\right)}^{\mathrm{Dynamic}}
    \label{eq:dynamical_contribution}
\end{equation}

Eq.~\ref{eq:dynamical_contribution} shows that relative changes in $P_{\mathrm{heavy}} $ are the sum of a well-understood, spatially-uniform ``thermodynamic'' increase in saturation specific humidity ($q_{\mathrm{sat}} $ -- see Eq.~\ref{eq:q_sat}), and a spatially-varying term. This spatially-varying term is the sum of $N $ regime probability shifts $\Delta\pi_{i}$ (changes in our unsupervised ML-derived convection cluster assignment probability or "cluster sizes" -- covered in more detail in Section~\ref{sec:Methodology}), and $N $ changes in regime characteristics $\Delta{\cal D}_{i}$ (changes in the ``dynamic contribution'' pre-factors, in precipitation units). 

Our simulation data already contain $P_{\mathrm{heavy}}$ and $q_{\mathrm{sat}} $, and we can derive $\pi_i$ from our unsupervised learning framework, giving us all the information we need to calculate the elusive pre-factors ${\cal D}_{i} $ and their changes with warming. Using equation \ref{eq:D}, we linearly regress $\frac{P_{\mathrm{heavy}}}{q_{\mathrm{sat}}} $ on the regime probabilities $\pi _i$ in both the reference and warm climates to derive the pre-factors ${\cal D}_{i} $, which are the weights of the multiple linear regression. This is a step toward understanding how the spatial patterns of storm-scale dynamical changes, which are notably hard to analyze, can affect the spatial patterns of heavy precipitation. Understanding these changes is critical to trust local climate change predictions.

\section{Machine Learning Approach \label{sec:Methodology}}

We will now discuss the data, models, and statistical techniques used in this paper. Additional details can be found in the Supplemental Information.

\subsection{Data: High-resolution, Earth-like Simulations of Global Surface Warming \label{sec:preprocessing}}

The multi-scale modeling framework~\citep{10.1175/BAMS-84-11-1547} used to generate our training and test data is composed of small, locally periodic 2D subdomains of explicit high-resolution physics that are embedded within each grid column of a coarser resolution ($1.9^{\circ} \times 2.5^{\circ}$ degree) host planetary circulation model~\cite{Khairoutdinov1999}. In total, we performed six simulations of present-day climate launched from different initial conditions (but consistent resolution), configured with storm resolving models that are 512 km in physical extent, each with 128 grid columns with a horizontal resolution of 4 km. We approximate the atmosphere with a simple bulk one-moment microphysical scheme and thirty vertical levels. We extract snapshots every four hours. We then perform six additional simulations but increase the sea surface temperatures by 4K. We compare the ``Control'' simulations against those with uniform increases in sea surface temperatures (``Warmed''). For our purposes, this creates a testbed for climate change, but we acknowledge that surface warming is only an approximation for the thermodynamic consequences of CO$_{2} $ concentration increase. 

To investigate the ``dynamic mode'' of precipitation, we choose vertical velocity to represent the state of the atmosphere. These vertical velocity fields contain information about complex updraft and gravity wave dynamics across multiple scales. We considered the entire 15S-15N latitude band containing diverse tropical convective regimes. Examples of these vertical velocity snapshots, selected by precipitation percentile, can be seen in Fig~\ref{fig:W_Fields}.

\subsection{VAE Training \label{sec:training}}

\begin{figure}
\centering
\includegraphics[width=0.5\textwidth]{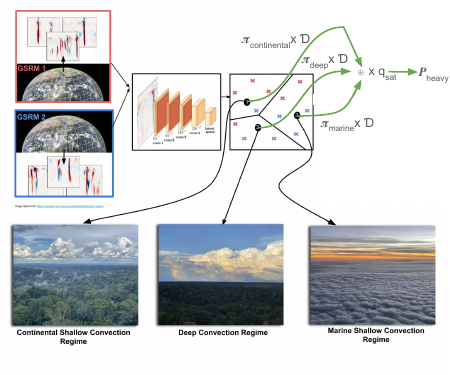}
\caption{\label{fig:VAE_Design} The VAE based approach to understand changes in heavy precipitation, $P_{heavy}$. Given a vertical velocity field w, the VAE non-linearly reduces the input dimension to yield a latent representation $\z$. We can cluster the simulated data sets (control and warmed) through their latent representations into three recognizable regimes of convection (continental shallow, deep, and marine shallow). The corresponding cluster assignment probabilities, $\pi$'s allow us to also quantify the dynamical contribution, $\cal D$ to heavy precipitation. Photos taken by Griffin Mooers.}
\end{figure}

Our ML methodology identifies dynamical regimes on the basis of two million two-dimensional vertical velocity fields. We first nonlinearly reduce the dimensionality of the data using a fully-convolutional VAE design, whose architecture is depicted in Fig~\ref{fig:VAE_Design}. To train the VAE, we perform beta-annealing~\cite{Higgins2017betaVAELB,Bowman2016GeneratingSF}, expanding the Evidence Lower Bound (ELBO) traditionally used to train the VAE by including a $\beta$ parameter and linearly anneal $\beta$ from 0 to one over 1600 training epochs. The number of layers and channels in the encoder and decoder are depicted in Fig~\ref{fig:VAE_Design} (4 layers in each, stride of two). After manual hyperparameter tuning, we choose ReLUs as activation functions in both the encoder and the decoder. We pick a relatively small kernel size of 3 to preserve the small-scale updrafts and downdrafts of our vertical velocity fields. The dimension of our latent space is 1000. 

We refer the reader to SI B for the details of the VAE's benchmarking and performance evaluation and proceed to use that VAE to investigate how convective regimes respond to climate change.

\subsection{Quantization Procedure}

The goal of the following analysis is to analyze upper quantile precipitation patterns arising from climate shifts, leveraging vertical velocity fields. The sheer amount of data requires additional statistical analysis. 

Although the use of a VAE encoder makes our high-resolution simulation data more manageable, we require additional work to derive the formal convective probability information, $ \pi $. The main idea is to convert a high-dimensional, continuous probability distribution over velocity fields into a fixed-size, discrete probability distribution over quantization points~\cite{Nas_1988}. Then, we use the coarse-grained, discrete distribution to compute various quantities of interest.

We use a convolutional VAE to nonlinearly embed our 2D input data (vertical velocity fields) into a lower-dimensional latent space. To quantize the emergent latent space, we employ k-means clustering (More details in SI-A): we encode our training data into the latent space and cluster them into  $N$ clusters. We also define a vector of \emph{cluster assignment probabilities} $\pi_i$ for $i=1, \cdots, N$ as the percentage of training data assigned to each cluster $i$. This dimensionality reduction and clustering can be thought of as a lossy compression of the data~\cite{yang2022introduction}. As we will see, the discrete structure helps us compute various quantities of interest. While the quantization approximation can, in principle, be made arbitrarily precise using a large $N$, we use $N=3$ in practice for interpretability based on the findings of~\cite{mooers2022comparing}. We cluster convection into three distinct regimes: (1) Marine Shallow Convection, (2) Continental Shallow Cumulus Convection, and (3) Deep Convection. We have found that higher counts of $N$ created repetitive regimes of Deep Convection while lower counts do not properly separate out these three distinct species of convection~\cite{mooers2022comparing}. 

We are especially interested in \emph{changes} of the cluster assignment probabilities under global warming. In order to compare the present and future data distributions, we train the VAE and learn the cluster centers based on present climate data. This yields the present cluster assignment probabilities $\pi^{0K}$. In a second step, we encode all future climate data into the latent space and assign each datum to the nearest (control) cluster center, yielding the future cluster assignment probabilities $\pi^{4K}$. The difference vector of assignment probabilities, before and after global warming, is given by $\Delta \pi = \pi^{4K} - \pi^{0K}$. This information can then be used as a proxy for dynamical regime shifts with warming. We visualize these shifts and interpret their implications for heavy precipitation below.

\section{Results\label{sec:results}}

\begin{figure}
\includegraphics[width=\linewidth]{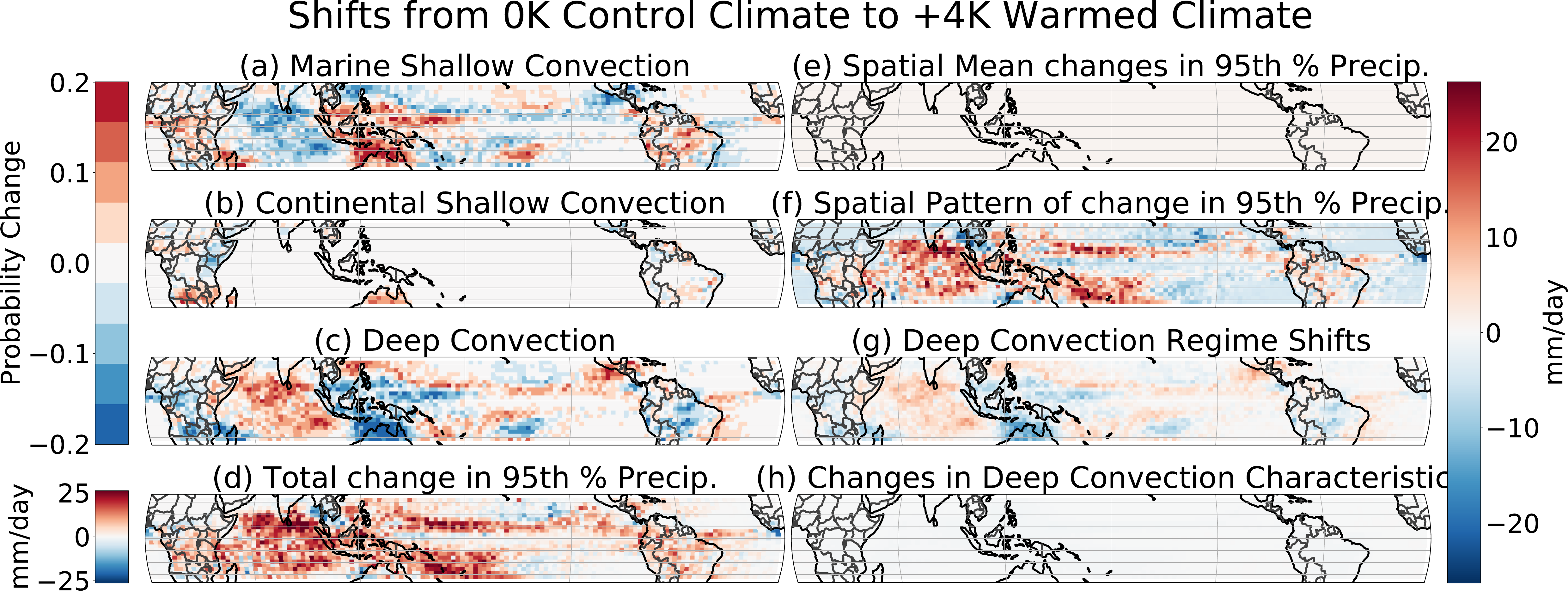}
\caption{\label{fig:Climate_Change} 
Changes induced by $+4^{\circ}$C of simulated global warming. Panels (a-c) display normalized probability shifts ($\Delta \pi$'s) in the three dynamical regimes found through clustering with $N=3$, corresponding to (a) ``marine shallow'', (b) ``continental shallow cumulus'', and (c) ``deep'' convection. The difference in the 95th percentile in precipitation between the control and warmed simulations are shown in (d). We subtract the spatial-mean change (e, the ``thermodynamics'' from Eq.~\ref{eq:q_sat}) from the total change (d) to yield the ``dynamic'' contribution (f). Using Eq.~\ref{eq:dynamical_contribution}, we decompose the changing spatial patterns (f) into five terms, including (g) probability changes in deep convection, (h) changes in deep convective precipitation, and three additional terms depicted in Fig. S5. 
\textbf{The patterns of storms change} (a-c), \textbf{which changes the patterns of heavy precipitation} (f), \textbf{mostly because deep convective storms shift location} (g).}
\end{figure}

\subsection{Unsupervised Machine Learning Reveals Convective Responses to Climate Change}

Fig~\ref{fig:Climate_Change} shows the probability shifts in convection type ($\Delta \pi_{i}$) from the ``Control'' to the ``Warmed'' climate. The dominant climate change signal captured by our unsupervised framework is the increased geographic concentration of deep convection (Fig~\ref{fig:Climate_Change}c). More specifically, deep convection becomes more probable over warm ocean waters and especially the Pacific Warm Pool~\cite{Adams_2014} while shallow convection becomes less common in these unstable regions (Fig~\ref{fig:Climate_Change}a). This result is consistent with observational trends showing an intensification of already powerful storms over the warm tropical waters~\cite{Adams_2014}. At first glance, the pattern of this unsupervised deep convection shift with warming ($\Delta \pi_1$) looks quite similar to shifts in upper precipitation quantiles (Fig~\ref{fig:Climate_Change}c vs. f).

With just the information from the regime probabilities, we can model the spatial patterns of precipitation changes at upper percentiles (Fig. S4). Our model becomes less accurate at lower precipitation quantiles, partly because we are not using specific humidity information (the approximation of Eq.~\ref{eq:Pextreme_w_qsat} is only valid for high precipitation quantiles). This degree of accuracy at the upper percentiles suggests that changes in the location of convective dynamical regimes can explain a large fraction of changes in heavy precipitation, which should be further tested in diverse climate change modeling frameworks. 

\subsection{Decomposing the Dynamic Contribution to Heavy Precipitation Changes}

We now isolate the dynamical contributions to dynamical changes in heavy precipitation by decomposing the spatial patterns (\ref{fig:Climate_Change}d) into changes in regime probability $\pi_i$ and changes in regime characteristics ${\cal D}_i$. Unlike traditional approaches that spatially average information, we use our fully-convolutional encoder and latent space clustering to leverage storm-scale variability.  

We calculated changes in regime probability (how regimes move in space) in section \ref{sec:Methodology}, so we must now calculate changes in how each regime produces precipitation, which involves the following two steps. First, we empirically estimate ${\cal D}_i$ by using the probabilities of deep and shallow convection\footnote{More specifically, we estimate the dynamical pre-factors (${\cal D}_i$) by regressing $P_{\mathrm{heavy}}/q_{\mathrm{sat}}$ on $\pi_1$ and $\pi_2$, neglecting the ``Continental Shallow Cumulus'' regime as it concentrates over arid continental zones with high lower tropospheric stability and low latent heat fluxes, making conditions unfavorable for precipitation~\cite{Dror_Part_3}.}. Second, we estimate changes in ``deep'' and ``shallow'' convection dynamical pre-factors as $\Delta {\cal D}_{i} = {\cal D}^{4K}_i -{\cal D}^{0K}_i$. 

\begin{figure}[ht!]
\centering
\includegraphics[height=8cm, width=0.75\linewidth]{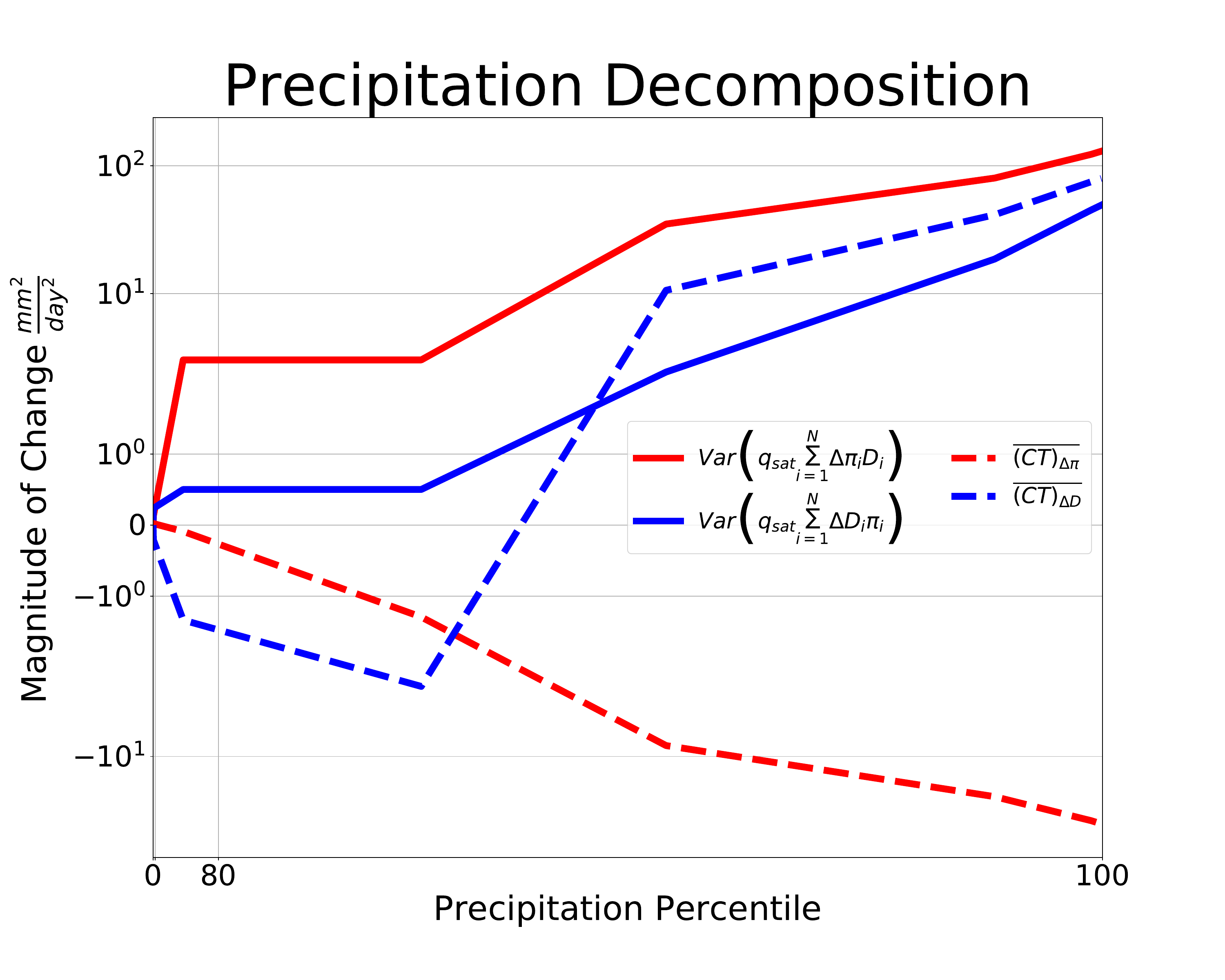}
\caption{\label{fig:Main_Decompostion} 
Based on Eq.~\ref{eq:spatial_anomaly} we decompose the total magnitude of the change in heavy precipitation. In our decomposition, we compare the mean of the spatial anomaly of convective probability shifts ($\Delta \pi$) to the changes in the dynamical prefactors ($\Delta D$). We find that the convective regime shifts are of greater importance to explain the changes in heavy precipitation (80th-99.99th percentiles)}
\end{figure}

We now have the requisite information to understand the drivers of heavy precipitation changes themselves. We ask: \textit{Did the patterns of heavy precipitation simply follow the changing patterns of the convective regime, or are there more complex changes in how deep convection produces rain?} We address this question by comparing how much of the spatial variance in heavy precipitation $\Delta P_{heavy}$ can be explained by changes in convection probability ($\Delta \pi$), and how much of it can be explained by changes in the dynamical prefactors ($\Delta {\cal D}$). This comparison relies on the following decomposition of heavy precipitation variance $\mathrm{var}\left(\Delta P_{\mathrm{heavy}}\right) $, derived in Sec~D of the SI:  

\begin{equation}
\mathrm{var}\left(\Delta P_{\mathrm{heavy}}\right) =\underbrace{\mathrm{var}\left[\left(q_{\mathrm{sat}}\sum_{i=1}^{N}\Delta\pi_{i}{\cal D}_{i}\right)\right]+\overline{\left(\mathrm{CT}\right)_{\Delta\pi}}}_{\mathrm{Shift\ in\ regime\ location}}+\underbrace{\mathrm{var}\left[\left(q_{\mathrm{sat}}\sum_{i=1}^{N}\pi_{i}\Delta{\cal D}_{i}\right)\right]+\overline{\left(\mathrm{CT}\right)_{\Delta{\cal D}}}}_{\mathrm{Intra-regime\ changes}}\ +\ {\cal R}
\label{eq:spatial_anomaly}
\end{equation}
where CT are cross-terms and $\cal{R}$ groups all terms of the decomposition that are not needed to compare differences in precipitation from regime shifts vs. intra-regime changes. To understand what is most crucial to precipitation changes, we depict the spatial mean of Eq.~\ref{eq:spatial_anomaly}'s terms in Fig~\ref{fig:Main_Decompostion}. 

For high precipitation quantiles where our model works best (especially above the 80th percentile), changes in the upper quantiles are dominated by regime probability shifts rather than by intra-regime changes (Fig~\ref{fig:Main_Decompostion}): The ``$\Delta {\cal D}$ term'' is smaller than the ``$\Delta \pi$ term'' (red line vs. blue line) \footnote{Note that at precipitation quantiles larger than 0.99, we lack samples for the analysis to work properly as evidenced by the pixelation of the changing patterns in Fig. S3}. This result aligns well with our qualitative analysis in Fig~\ref{fig:Climate_Change} showing the similarity between the spatial patterns of deep convection regime changes and heavy precipitation changes. The spatial patterns of heavy precipitation changes are dominated by the changing patterns of storm characteristics identified by our unsupervised framework while changes in how each regime produces precipitation are less important.

We are also confident in the robustness of these results based on the degree of similarity in the findings of the importance of deep convection shifts in both our work and~\cite{Tan_Nature}. The fact that in satellite and reanalysis data probability shifts in convection type are also the drivers of changes in extreme precipitation is a reassuring measure of support for our findings as well as for the performance of the GSRM representation of the global high-resolution vertical velocity structure more broadly. At the same time, it suggests that machine learning approaches like the one we deploy with minimal domain knowledge are just as faithful to the physics of the atmosphere as more traditional convection analysis.

\section{Conclusion}

Based on our findings, proper prediction of spatial shifts in deep convection with global warming should allow us to anticipate regional and local changes in heavy precipitation. This highlights the importance of leveraging the full spatial extent of this information (traditionally averaged out) to derive accurate regional and local changes. Although our findings were achieved from unsupervised learning, we are reassured that similar findings can be derived in observational satellite data from regime-based analysis of deep convection as well~\cite{Tan_Nature}. The necessity of understanding this rich spatial information indicates a role for ML methods like variational encoders and clustering routines for the analysis of storm-scale climate information to deepen our understanding of intense events. Our next step to build credibility in this unsupervised model is to deploy the workflow on more diverse climate change data like the High-Resolution Model Inter-comparison Project (HighResMIP)~\cite{Eyring_2016, Haarsma_2016} and determine its ability to explain spatial variations in heavy precipitation with climate change.

\begin{Backmatter}

\paragraph{Acknowledgments}

The authors thank Paul O'Gorman, Harshini Mangipudi, Pierre Gentine, Veronika Eyring, Prakhar Srivastava, and Liran Peng for useful conversations that helped advance this work.

\paragraph{Funding Statement}

The authors thank the Department of Energy under grant DE-SC0022331, the National Science Foundation (NSF) Machine Learning and Physical Sciences (MAPS) program and NSF grant 1633631, Office of Advanced Cyberinfrastructure grant OAC-1835863, Division of Atmospheric and Geospace Sciences grant AGS-1912134, Division of Information and Intelligent Systems grants IIS-2047418, IIS-2003237, IIS-2007719, Division of Social and Economic Sciences grant SES-1928718, and Division of Computer and Network Systems grant CNS-2003237 for funding support and co-funding by the Enabling Aerosol-cloud interactions at GLobal convection-permitting scalES (EAGLES) project (74358), of the U.S. Department of Energy Office of Biological and Environmental Research, Earth System Model Development program area. We further acknowledge funding from NSF Science and Technology Center LEAP (Launching Early-Career Academic Pathways) award 2019625. Computational resources were provided by the Extreme Science and Engineering Discovery Environment supported by NSF Division of Advanced Cyberinfrastructure Grant number ACI-1548562 (charge number TG-ATM190002). DYAMOND data management was provided by the German Climate Computing Center (DKRZ) and supported through the projects ESiWACE and ESiWACE2. The projects ESiWACE and ESiWACE2 have received funding from the European Union's Horizon 2020 research and innovation programme under grant agreements No 675191 and 823988. This work was also supported by gifts from Disney and Qualcomm.

\paragraph{Competing Interests}

The authors have no conflicts of interest to declare.

\paragraph{Data Availability Statement}

The codebase for running the ``SPCAM5'' simulations is the same employed by~\cite{Parishani_Pritchard}, which is archived at~\url{https://github.com/mspritch/UltraCAM-spcam2_0_cesm1_1_1}; this code was in turn forked from a development version of the CESM1.1.1 located on the NCAR central subversion repository under tag~\url{spcam_cam5_2_00_forCESM1_1_1Rel_V09}, which dates to February 25, 2013. 

Training and postprocessing scripts, as well as saved model weights and Python environments, are available on GitHub at~\href{https://zenodo.org/badge/latestdoi/608769237}{10.5281/zenodo.7693052}.

\paragraph{Ethical Standards}

The research meets all ethical guidelines.

\paragraph{Author Contributions}

Data curation and visualizations: GM and TB,
Methodology: GM and TB,
Conceptualization: GM, TB, SM, MP,
Writing Manuscript: GM, TB, SM, MP

\bibliographystyle{abbrv}
\bibliography{main}

\begin{thebibliography}{10}

\bibitem{abbott2020convective}
T.~H. Abbott, T.~W. Cronin, and T.~Beucler.
\newblock Convective dynamics and the response of precipitation extremes to warming in radiative--convective equilibrium.
\newblock {\em Journal of the Atmospheric Sciences}, 77(5):1637--1660, 2020.

\bibitem{Adger_2007}
W.~Adger, S.~Agrawala, M.~Mirza, C.~Conde, K.~O'Brien, J.~Pulhin, R.~Pulwarty, B.~Smit, and K.~Takahashi.
\newblock Assessment of adaptation practices, options, constraints and capacity. climate change 2007: impacts, adaptation and vulnerability.
\newblock {\em Contribution of working group II to the fourth assessment report of the intergovernmental panel on climate change}, pages 717--743, 01 2007.

\bibitem{Adger_2009}
W.~N. Adger, S.~Dessai, M.~Goulden, M.~Hulme, I.~Lorenzoni, D.~R. Nelson, L.~O. Naess, J.~Wolf, and A.~Wreford.
\newblock Are there social limits to adaptation to climate change?
\newblock {\em Climatic Change}, 93(3):335--354, 2009.

\bibitem{Adams_2014}
R.~P. Allan, C.~Liu, M.~Zahn, D.~A. Lavers, E.~Koukouvagias, and A.~Bodas-Salcedo.
\newblock Physically consistent responses of the global atmospheric hydrological cycle in models and observations.
\newblock {\em Surveys in Geophysics}, 35(3):533--552, 2014.

\bibitem{Arthur_2007}
D.~Arthur and S.~Vassilvitskii.
\newblock K-means++: The advantages of careful seeding.
\newblock In {\em Proceedings of the Eighteenth Annual ACM-SIAM Symposium on Discrete Algorithms}, SODA '07, page 1027–1035, USA, 2007. Society for Industrial and Applied Mathematics.

\bibitem{Bowman2016GeneratingSF}
S.~R. Bowman, L.~Vilnis, O.~Vinyals, A.~M. Dai, R.~J{\'o}zefowicz, and S.~Bengio.
\newblock Generating sentences from a continuous space.
\newblock In {\em CoNLL}, 2016.

\bibitem{Davies_1979}
D.~L. Davies and D.~W. Bouldin.
\newblock A cluster separation measure.
\newblock {\em IEEE Transactions on Pattern Analysis and Machine Intelligence}, PAMI-1(2):224--227, 1979.

\bibitem{Denby_2020}
L.~Denby.
\newblock Discovering the importance of mesoscale cloud organization through unsupervised classification.
\newblock {\em Geophysical Research Letters}, 47(1):e2019GL085190, 2020.
\newblock e2019GL085190 10.1029/2019GL085190.

\bibitem{Dror_Part_3}
T.~Dror, V.~Silverman, O.~Altaratz, M.~D. Chekroun, and I.~Koren.
\newblock Uncovering the large-scale meteorology that drives continental, shallow, green cumulus through supervised classification.
\newblock {\em Geophysical Research Letters}, 49(8):e2021GL096684, 2022.
\newblock e2021GL096684 2021GL096684.

\bibitem{edelman2014state}
A.~Edelman, A.~Gedling, E.~Konovalov, R.~McComiskie, A.~Penny, N.~Roberts, S.~Templeman, D.~Trewin, and M.~Ziembicki.
\newblock {\em State of the Tropics - 2014 Report}.
\newblock 06 2014.

\bibitem{douville2021water}
A.~Edelman, A.~Gelding, E.~Konovalov, R.~McComiskie, A.~Penny, N.~E. Roberts, S.~Templeman, D.~Trewin, M.~Ziembicki, B.~Trewin, R.~Cortlet, J.~Hemingway, J.~L. Isaac, and S.~M. Turton.
\newblock State of the tropics 2014 report.
\newblock 2014.

\bibitem{emori2005dynamic}
S.~Emori and S.~Brown.
\newblock Dynamic and thermodynamic changes in mean and extreme precipitation under changed climate.
\newblock {\em Geophysical Research Letters}, 32(17), 2005.

\bibitem{Eyring_2016}
V.~Eyring, S.~Bony, G.~A. Meehl, C.~A. Senior, B.~Stevens, R.~J. Stouffer, and K.~E. Taylor.
\newblock Overview of the coupled model intercomparison project phase 6 (cmip6) experimental design and organization.
\newblock {\em Geoscientific Model Development}, 9(5):1937--1958, 2016.

\bibitem{gray1984vector}
R.~Gray.
\newblock Vector quantization.
\newblock {\em IEEE Assp Magazine}, 1(2):4--29, 1984.

\bibitem{Haarsma_2016}
R.~J. Haarsma, M.~J. Roberts, P.~L. Vidale, C.~A. Senior, A.~Bellucci, Q.~Bao, P.~Chang, S.~Corti, N.~S. Fu\v{c}kar, V.~Guemas, J.~von Hardenberg, W.~Hazeleger, C.~Kodama, T.~Koenigk, L.~R. Leung, J.~Lu, J.-J. Luo, J.~Mao, M.~S. Mizielinski, R.~Mizuta, P.~Nobre, M.~Satoh, E.~Scoccimarro, T.~Semmler, J.~Small, and J.-S. von Storch.
\newblock High resolution model intercomparison project (highresmip~v1.0) for cmip6.
\newblock {\em Geoscientific Model Development}, 9(11):4185--4208, 2016.

\bibitem{hess-22-2041-2018}
S.~Hettiarachchi, C.~Wasko, and A.~Sharma.
\newblock Increase in flood risk resulting from climate change in a developed urban watershed - the role of storm temporal patterns.
\newblock {\em Hydrology and Earth System Sciences}, 22(3):2041--2056, 2018.

\bibitem{Higgins2017betaVAELB}
I.~Higgins, L.~Matthey, A.~Pal, C.~Burgess, X.~Glorot, M.~M. Botvinick, S.~Mohamed, and A.~Lerchner.
\newblock beta-vae: Learning basic visual concepts with a constrained variational framework.
\newblock In {\em ICLR}, 2017.

\bibitem{10.1175/1520-0469(2003)060<0607:CRMOTA>2.0.CO;2}
M.~Khairoutdinov and D.~Randall.
\newblock Cloud resolving modeling of the arm summer 1997 iop: Model formulation, results, uncertainties, and sensitivities.
\newblock {\em Journal of The Atmospheric Sciences - J ATMOS SCI}, 60:607--625, 02 2003.

\bibitem{Khairoutdinov1999}
M.~F. Khairoutdinov and Y.~L. Kogan.
\newblock A large eddy simulation model with explicit microphysics: Validation against aircraft observations of a stratocumulus-topped boundary layer.
\newblock {\em Journal of the Atmospheric Sciences}, 56(13):2115 -- 2131, 1999.

\bibitem{kurihana2021datadriven}
T.~Kurihana, E.~Moyer, R.~Willett, D.~Gilton, and I.~Foster.
\newblock Data-driven cloud clustering via a rotationally invariant autoencoder, 2021.

\bibitem{li2020response}
Z.~Li and P.~A. O’Gorman.
\newblock Response of vertical velocities in extratropical precipitation extremes to climate change.
\newblock {\em Journal of Climate}, 33(16):7125--7139, 2020.

\bibitem{Lloyd_1957}
S.~Lloyd.
\newblock Least squares quantization in pcm.
\newblock {\em IEEE Transactions on Information Theory}, 28(2):129--137, 1982.

\bibitem{Macqueen67somemethods}
J.~Macqueen.
\newblock Some methods for classification and analysis of multivariate observations.
\newblock In {\em In 5-th Berkeley Symposium on Mathematical Statistics and Probability}, pages 281--297, 1967.

\bibitem{mooers2022comparing}
G.~Mooers, M.~Pritchard, T.~Beucler, P.~Srivastava, H.~Mangipudi, L.~Peng, P.~Gentine, and S.~Mandt.
\newblock Comparing storm resolving models and climates via unsupervised machine learning.
\newblock {\em arXiv preprint arXiv:2208.11843}, 2022.

\bibitem{10.1145/3429309.3429324}
G.~Mooers, J.~Tuyls, S.~Mandt, M.~Pritchard, and T.~G. Beucler.
\newblock Generative modeling of atmospheric convection.
\newblock In {\em Proceedings of the 10th International Conference on Climate Informatics}, CI2020, page 98–105, New York, NY, USA, 2021. Association for Computing Machinery.

\bibitem{muller2020response}
C.~Muller and Y.~Takayabu.
\newblock Response of precipitation extremes to warming: what have we learned from theory and idealized cloud-resolving simulations, and what remains to be learned?
\newblock {\em Environmental Research Letters}, 15(3):035001, 2020.

\bibitem{muller2011intensification}
C.~J. Muller, P.~A. O’Gorman, and L.~E. Back.
\newblock Intensification of precipitation extremes with warming in a cloud-resolving model.
\newblock {\em Journal of Climate}, 24(11):2784--2800, 2011.

\bibitem{Nas_1988}
N.~Nasrabadi and R.~King.
\newblock Image coding using vector quantization: a review.
\newblock {\em IEEE Transactions on Communications}, 36(8):957--971, 1988.

\bibitem{o2015precipitation}
P.~A. O’Gorman.
\newblock Precipitation extremes under climate change.
\newblock {\em Current climate change reports}, 1(2):49--59, 2015.

\bibitem{o2009scaling}
P.~A. O’gorman and T.~Schneider.
\newblock Scaling of precipitation extremes over a wide range of climates simulated with an idealized gcm.
\newblock {\em Journal of Climate}, 22(21):5676--5685, 2009.

\bibitem{Parishani_Pritchard}
H.~Parishani, M.~S. Pritchard, C.~S. Bretherton, M.~C. Wyant, and M.~Khairoutdinov.
\newblock Toward low-cloud-permitting cloud superparameterization with explicit boundary layer turbulence.
\newblock {\em Journal of Advances in Modeling Earth Systems}, 9(3):1542--1571, 2017.

\bibitem{pendergrass2014changes}
A.~G. Pendergrass and D.~L. Hartmann.
\newblock Changes in the distribution of rain frequency and intensity in response to global warming.
\newblock {\em Journal of Climate}, 27(22):8372--8383, 2014.

\bibitem{10.1175/BAMS-84-11-1547}
D.~Randall, M.~Khairoutdinov, A.~Arakawa, and W.~Grabowski.
\newblock {Breaking the Cloud Parameterization Deadlock}.
\newblock {\em Bulletin of the American Meteorological Society}, 84(11):1547--1564, 11 2003.

\bibitem{ROUSSEEUW198753}
P.~J. Rousseeuw.
\newblock Silhouettes: A graphical aid to the interpretation and validation of cluster analysis.
\newblock {\em Journal of Computational and Applied Mathematics}, 20:53--65, 1987.

\bibitem{Tan_Nature}
J.~Tan, C.~Jakob, W.~B. Rossow, and G.~Tselioudis.
\newblock Increases in tropical rainfall driven by changes in frequency of organized deep convection.
\newblock {\em Nature}, 519(7544):451--454, 2015.

\bibitem{Wang_2009}
Z.~Wang and A.~Bovik.
\newblock Bovik, a.c.: Mean squared error: love it or leave it? - a new look at signal fidelity measures. ieee sig. process. mag. 26, 98-117.
\newblock {\em Signal Processing Magazine, IEEE}, 26:98 -- 117, 02 2009.

\bibitem{yang2022introduction}
Y.~Yang, S.~Mandt, and L.~Theis.
\newblock An introduction to neural data compression.
\newblock {\em arXiv preprint arXiv:2202.06533}, 2022.

\end{thebibliography}

\section*{Supplemental Information}

We wish to provide additional details about our decomposition discussed in Section 2 and the choices and performance behind our unsupervised machine learning framework from Section 3. We start by discussing information on the hyper-parameters choices for the clustering that leads to our dynamical regime shifts, $\pi$'s. As a justification for our use of a Variational Autoencoder, we provide a comparison of the performance against a simpler linear baseline. Finally, we also include additional steps showing how we decompose the full spatial variance of heavy precipitation for analysis as well as plotting all the terms of the full decomposition. 

\subsection*{A. K-Means Clustering Approach}
\label{sec:k_means_details}
We apply the K-Means Clustering algorithm to partition the latent space of our VAE and analyze which physical properties can be clustered in this reduced order $\z$ space. This approach first randomly assigns centroids, C, to locations in the $\z$ space (note we actually use the more modern k++ algorithm~\cite{Arthur_2007} to maximize the initial distances between the centroids). Latent representations of each sample $\z_i$, in the test dataset of size $N$, are assigned to their nearest centroid. The second stage of the algorithm moves the centroid to the mean of the assigned cluster. The process repeats until the sum of the square distances (or the Inertia, $I$) between the latent space data points and the centroids are minimized~\cite{Lloyd_1957, Macqueen67somemethods} such that:

 \begin{equation}
     \mathrm{\overline{I} \overset{\mathrm{def}}{=}\sum_{i=0}^{N}\min_{l \in C}{||z_i - \bar{z}_l||^2}}, 
     \label{eq:kmeans}
 \end{equation}
in which $\bar{z}_l$ is the mean of the given samples belonging to a cluster l for the total number of cluster centers C. We always calculate ten different K-means initializations and then select the initialization with the lowest inertia. This process allows us to derive the three data-driven convection regimes within SPCAM highlighted in Fig~\ref{fig:Climate_Change}.

We qualitatively choose an optimal number of cluster centroids (centers), $k$ by incorporating domain knowledge rather than a traditional approach relying on the rate of decrease in $I$ as $k$ increases or a single quantitative value such as a Silhouette Coefficient~\cite{ROUSSEEUW198753} or Davies-Bouldin Index~\cite{Davies_1979}. More specifically, we identify the maximum number of ``unique clusters''. We define a ``unique cluster'' of convection as a group in the latent space where the typical physical properties (vertical structure, intensity, and geographic domain) of the vertical velocity fields are not similar to the physical properties of another group elsewhere in the latent space. Empirically this exercise enables us to create three unique regimes of convection (Fig~\ref{fig:Climate_Change}). When we increase $k$ above three, we get sub-groups of ``Deep Convection'' without differences in either vertical mode, intensity, or geography. Thus we don't consider $N > 3$ to be physically meaningful for our purposes. 

 Because we seek to contrast common clusters between different climates, we do not use Agglomerative (hierarchical) Clustering unlike other recent works that cluster compressed representations of clouds from ML models~\cite{Denby_2020, kurihana2021datadriven}. Using the K-means approach, we can save the cluster centroids at the end of the algorithm. This provides a basis for cluster assignments for latent representations of out-of-sample test datasets when we use a common encoder as in Section B. More specifically, we only use the cluster centroids to get label assignments in other latent representations. We don't move the cluster centroids themselves once they have been optimized on the original test dataset (the second part of the K-means algorithm). Keeping the center of the clusters the same between different types of test data ensures we can objectively contrast cluster differences through the lens of the common latent space.

\subsection*{B. VAE Benchmarking and Performance Evaluation \label{sec:baselines}}

We train our VAE on 160,000 unique vertical velocity fields and use an additional 125,000 samples to validate and optimize the model hyperparameters. Finally, we leverage 1,000,000 vertical velocity fields in the test dataset for robust analysis. The high count in the test dataset is necessary both due to the high spatio-temporal correlations common in meteorological data but also because of the geographic conditioning in our analysis -- we need enough samples at each lat/lon grid cell, not just globally. To determine whether our data are nonlinear enough to warrant the use of a VAE we also train a baseline model of the same architecture but with all activation functions replaced by ``linear''. The fact that the VAE reconstructs the vertical velocity snapshots with both lower error and a higher degree of structural similarity suggests significant non-linearity is involved in compressing and rebuilding the 2D fields (Tab~\ref{tab:MSE_Table} and Tab~\ref{tab:SSIM_Table}). This problem is therefore well suited for the non-linear dimensionality reduction of the VAE encoder and less so for linear models.

\begin{table}
\centering
\begin{tabular}{ |c|c|c|c| }
\hline
\multicolumn{4}{|c|}{Mean Squared Error $m^2/s^2$} \\
\hline
Model & Training Set & Validation Set & Test Set  \\
\hline
VAE & $3.79 * 10^{-4}$ & $1.11 * 10^{-3}$ &  $3.33*10^{-3}$  \\
Linear Baseline  & $3.10*10^{-3}$ & $4.70*10^{-3}$ & $5.10*10^{-2}$ \\
\hline
\end{tabular}
\caption{The MSE of both of our models (``linear baseline'' and VAE) calculated across training/validation/test data. For both training and test data, we see low reconstruction errors, suggesting satisfactory skill and generalization ability. Overall, the VAE outperforms the ``linear'' baseline}
\label{tab:MSE_Table}
\end{table}

\begin{table}
\centering
\begin{tabular}{ |c|c|c|c| }
\hline
\multicolumn{4}{|c|}{Structural Similarity Index Metric} \\
\hline
Model & Training Set & Validation Set & Test Set \\
\hline
VAE & $0.998$ & $0.995$  & $0.987$ \\
Linear Baseline & $0.990$ & $0.986$ & $0.981$ \\
\hline
\end{tabular}
\caption{The mean SSIM~\cite{Wang_2009} of both of our models across training/validation/test data. The models both generalize well to our test data. Again, the VAE outperforms the ``linear baseline''}
\label{tab:SSIM_Table}
\end{table}

Tab~\ref{tab:MSE_Table} and Tab~\ref{tab:SSIM_Table} show 160,000 is enough training samples to create reconstructions of high-resolution vertical velocity fields with both a low MSE and a high degree of overall structural similarity. Though there is a small amount of overfitting, we see that performance remains strong for a test dataset containing multiple species of convection from all parts of the tropics ranging from deserts to rainforests; oceans to continents. 

\renewcommand{\thefigure}{}
\captionsetup[figure]{name={Figure S1}}
\begin{figure}[ht!]
\centering
\includegraphics[height=8cm, width=0.75\linewidth]{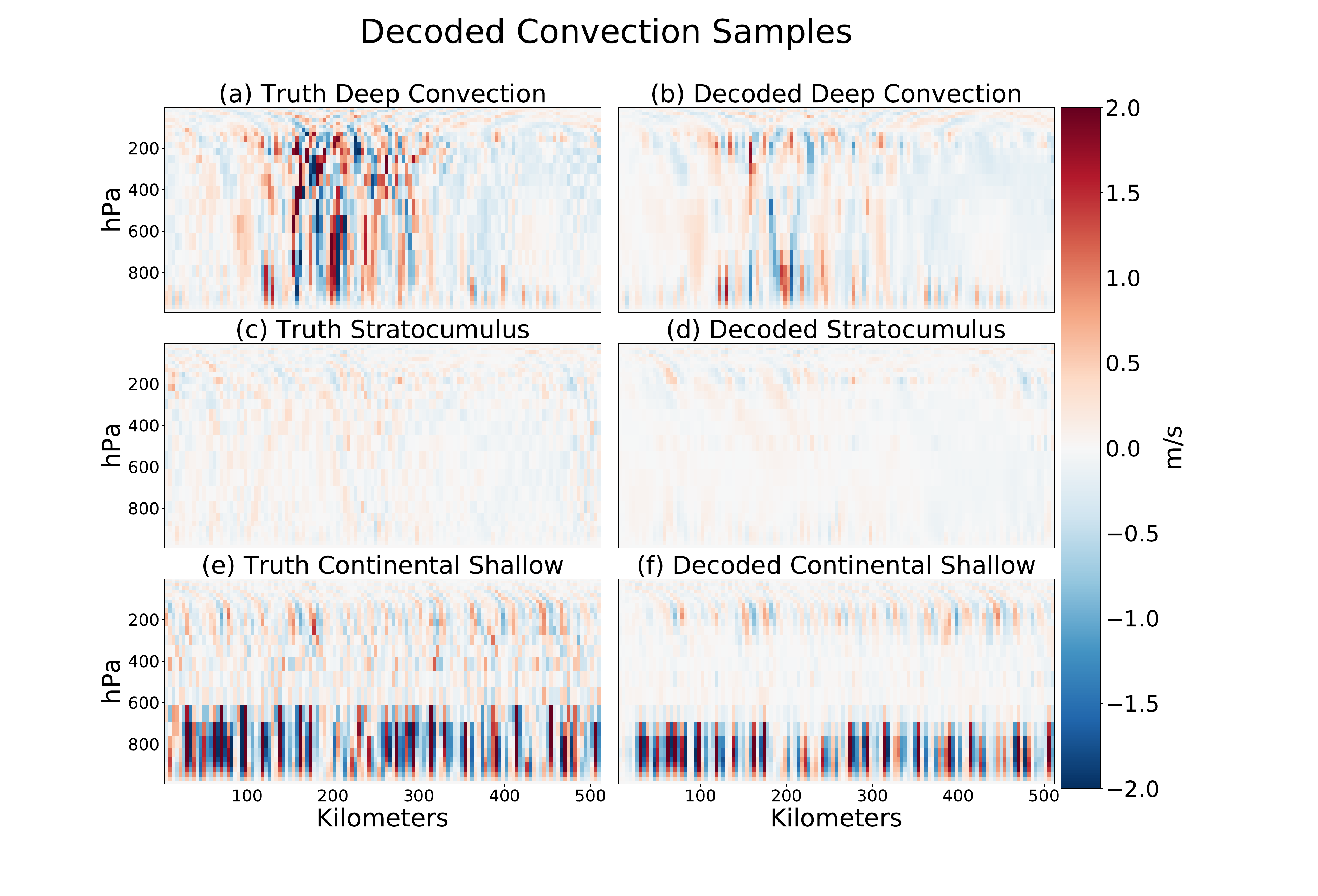}
\caption{\label{fig:Reconstructions} 
We plot the sample nearest to the centroid in the latent space of each of our three regimes of convection before being passed through the VAE encoder (a,c,e) and what those convection samples look like after being decoded by the VAE (b,d,f). We use this as a proxy for the centroids themselves, which at the high dimension of our latent space (1000) are themselves on a different manifold. Through this proxy, we can gleam the nature of the convection in each type of regime and how the VAE latent space organizes itself. Qualitatively, we see the VAE is able to reconstruct different forms of convection across different spatial scales.}
\end{figure}

Qualitatively, we also show decoded snapshots nearest the centroids from each of our three regimes of tropical convection. Overall reconstruction quality is good and we can see spatial structures from a variety of scales reproduced through the decoder of the VAE (Fig. S1 a,c,e vs. b,d,f). There is some under-prediction of the most intense up and downdrafts of convection (Fig. S1b), but this is to be expected as our loss is optimized to promote disentanglement of the latent space over reconstruction accuracy. If our goal was reconstruction accuracy instead of interpretability, we could add a statistical constraint to the loss function as we did in~\cite{10.1145/3429309.3429324} to better reconstruct extremes and small-scale variance.

\subsection*{C. Full Decomposition of the Spatial Variance of Heavy Precipitation}\label{sec:precip_eqs}

We derive the decomposition of the spatial variance of heavy precipitation in four steps. First, we multiply Eq. (2.7) by $P_{\mathrm{heavy}}$:
\begin{equation}
\Delta P_{\mathrm{heavy}}=\underbrace{P_{\mathrm{heavy}}\frac{\Delta q_{\mathrm{sat}}}{q_{\mathrm{sat}}}}_{{\cal Q}}+q_{\mathrm{sat}}\left(\Delta{\cal D}_{0}+\sum_{i=1}^{N}\Delta\pi_{i}{\cal D}_{i}+\sum_{i=1}^{N}\pi_{i}\Delta{\cal D}_{i}\right).
\label{eq:27_multiplied}
\end{equation}
For convenience, we use ${\cal Q}$ to denote the first term of Eq.~\ref{eq:27_multiplied}. We then take its spatial anomaly by applying the spatial anomaly operator $\left(X^{\prime}\right)$:
\begin{equation}
\Delta P_{\mathrm{heavy}}^{\prime}={\cal Q}^{\prime}+\Delta{\cal D}_{0}q_{\mathrm{sat}}^{\prime}+\left(q_{\mathrm{sat}}\sum_{i=1}^{N}\Delta\pi_{i}{\cal D}_{i}\right)^{\prime}+\left(q_{\mathrm{sat}}\sum_{i=1}^{N}\pi_{i}\Delta{\cal D}_{i}\right)^{\prime},
\label{eq:spatial_anomaly_equation}
\end{equation}
where we have used the fact that $\Delta{\cal D}_{0}\ $is uniform in space ($\Delta{\cal D}_{0}=\overline{\Delta{\cal D}_{0}}$ and $\Delta{\cal {\cal D}}_{0}^{\prime}=0$). Squaring Eq~\ref{eq:spatial_anomaly_equation} yields:
\begin{equation}
\left(\Delta P_{\mathrm{heavy}}^{\prime}\right)^{2}=\left({\cal Q}^{\prime}\right)^{2}+\left(\Delta{\cal D}_{0}\right)^{2}\left(q_{\mathrm{sat}}^{\prime}\right)^{2}+\left[\left(q_{\mathrm{sat}}\sum_{i=1}^{N}\Delta\pi_{i}{\cal D}_{i}\right)^{\prime}\right]^{2}+\left[\left(q_{\mathrm{sat}}\sum_{i=1}^{N}\pi_{i}\Delta{\cal D}_{i}\right)^{\prime}\right]^{2}+\mathrm{CT},
\label{eq:square_anomaly}
\end{equation}
where the cross-terms CT can be decomposed into cross-terms involving spatial shifts in regime probability:
\begin{equation}
\left(\mathrm{CT}\right)_{\Delta\pi}\overset{\mathrm{def}}{=}2\left(q_{\mathrm{sat}}\sum_{i=1}^{N}\Delta\pi_{i}{\cal D}_{i}\right)^{\prime}\left[{\cal Q}^{\prime}+\Delta{\cal D}_{0}q_{\mathrm{sat}}^{\prime}\right],
\end{equation}
cross-terms involving changes in how each regime produces precipitation:
\begin{equation}
\left(\mathrm{CT}\right)_{\Delta{\cal D}}\overset{\mathrm{def}}{=}2\left(q_{\mathrm{sat}}\sum_{i=1}^{N}\pi_{i}\Delta{\cal D}_{i}\right)^{\prime}\left[{\cal Q}^{\prime}+\Delta{\cal D}_{0}q_{\mathrm{sat}}^{\prime}\right],
\end{equation}

and additional cross-terms:
\begin{equation}
\left(\mathrm{CT}\right)_{\mathrm{other}}\overset{\mathrm{def}}{=}2\Delta{\cal D}_{0}q_{\mathrm{sat}}^{\prime}{\cal Q}^{\prime}+2\left(q_{\mathrm{sat}}\sum_{i=1}^{N}\Delta\pi_{i}{\cal D}_{i}\right)^{\prime}\left(q_{\mathrm{sat}}\sum_{i=1}^{N}\pi_{i}\Delta{\cal D}_{i}\right)^{\prime}
\end{equation}
Taking the spatial mean $\left(\overline{X}\right)$ of Eq~\ref{eq:square_anomaly} and noting that the spatial variance is defined as the spatial mean of the squared spatial anomaly, we derive the following decomposition:
\begin{equation}
\begin{aligned}\mathrm{var}\left(\Delta P_{\mathrm{heavy}}\right) & =\mathrm{var}\left({\cal Q}\right)+\left(\Delta{\cal D}_{0}\right)^{2}\mathrm{var}\left(q_{\mathrm{sat}}\right)+\mathrm{var}\left[\left(q_{\mathrm{sat}}\sum_{i=1}^{N}\Delta\pi_{i}{\cal D}_{i}\right)\right]+\mathrm{var}\left[\left(q_{\mathrm{sat}}\sum_{i=1}^{N}\pi_{i}\Delta{\cal D}_{i}\right)\right]\\
 & +\overline{\left(\mathrm{CT}\right)_{\Delta\pi}}+\overline{\left(\mathrm{CT}\right)_{\Delta{\cal D}}}+\overline{\left(\mathrm{CT}\right)_{\mathrm{other}}}+\overline{\mathrm{Numerical\ Residual}},
\end{aligned}
\label{eq:full_eq_decomposition}
\end{equation}
where we have introduced the decomposition's numerical residual, which helps us assess which terms are significant. Grouping the terms irrelevant to the comparison between regime spatial shifts and intra-regime changes into a single term, ${\cal R}$, mathematically defined as:
\begin{equation}
{\cal R}\overset{\mathrm{def}}{=}\mathrm{var}\left({\cal Q}\right)+\left(\Delta{\cal D}_{0}\right)^{2}\mathrm{var}\left(q_{\mathrm{sat}}\right)+\overline{\left(\mathrm{CT}\right)_{\mathrm{other}}}+\overline{\mathrm{Numerical\ Residual}},
\end{equation}
we recover Eq. (4.1) from the manuscript's main text. For additional context on the significance of our decomposition, we plot all the terms in Fig. S2. We see that as in Fig. S2, our decomposition is most valid for high precipitation percentiles (percentiles where the residual (blue line) is of lesser magnitude than other quantities).

\captionsetup[figure]{name={Figure S2}}
\begin{figure}[ht!]
\centering
\includegraphics[height=8cm, width=0.75\linewidth]{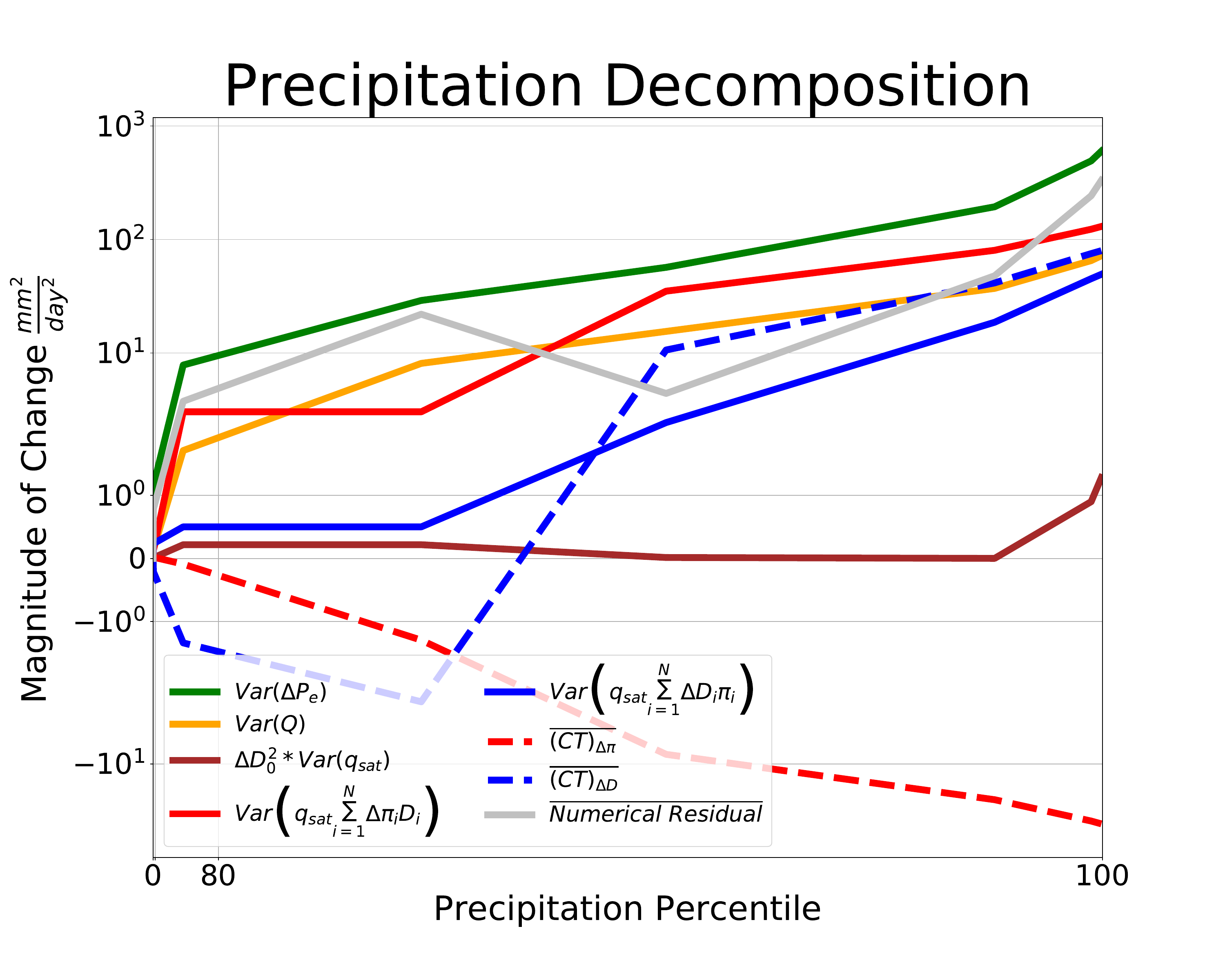}
\caption{\label{fig:SI_Full_Decompostion} 
Derived from Eq.~\ref{eq:full_eq_decomposition}, we plot each term from the full decomposition for the variance in the change in heavy precipitation, $Var (\Delta P_{e}^{^2})$. We focus primarily on precipitation percentiles 80-99, where our model is valid (the numerical residual, grey, is smaller than the key terms) and we have sufficient data (Fig. 3). Across these upper quantiles of precipitation, we find that the change in probability of convection type ($\Delta \pi$ -- red) is of greater importance than changes in the Dynamical Prefactors ($\Delta D$ -- blue). For additional context compared to Figure~\ref{fig:Main_Decompostion}, we include all terms from Eq.~\ref{eq:full_eq_decomposition}}
\end{figure}

\subsection*{D. Supplemental Figures}

\captionsetup[figure]{name={Figure S3}}
\begin{figure}[ht!]
\centering
\includegraphics[width=0.75\linewidth]{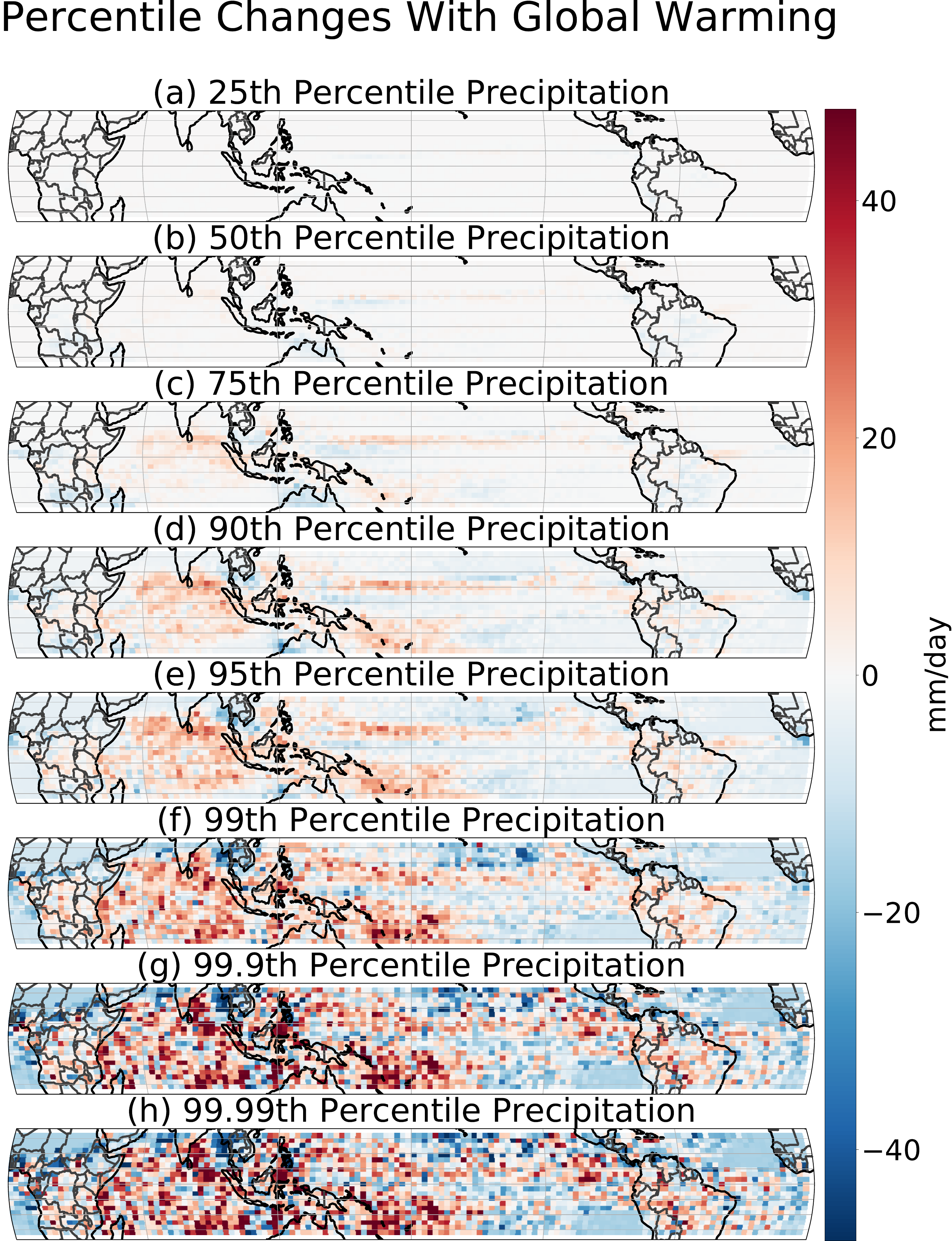}
\caption{\label{fig:extra_precip_percentiles}
The shifts in different percentiles of precipitation with global warming, where we again stratified and plotted the data by latitude/longitude grid cell. As in Fig~\ref{fig:Climate_Change}d we again remove the mean to highlight the dynamical pattern and see at what threshold the alignment with the VAE identified Deep Convection shifts (Fig~\ref{fig:Climate_Change}c) is greatest. The top percentiles including (f-h) are pixelated because of a lack of samples that are out on the tail of the PDF.}
\end{figure}

\captionsetup[figure]{name={Figure S4}}
\begin{figure}[ht!]
\centering
\includegraphics[height=8cm, width=0.65\linewidth]{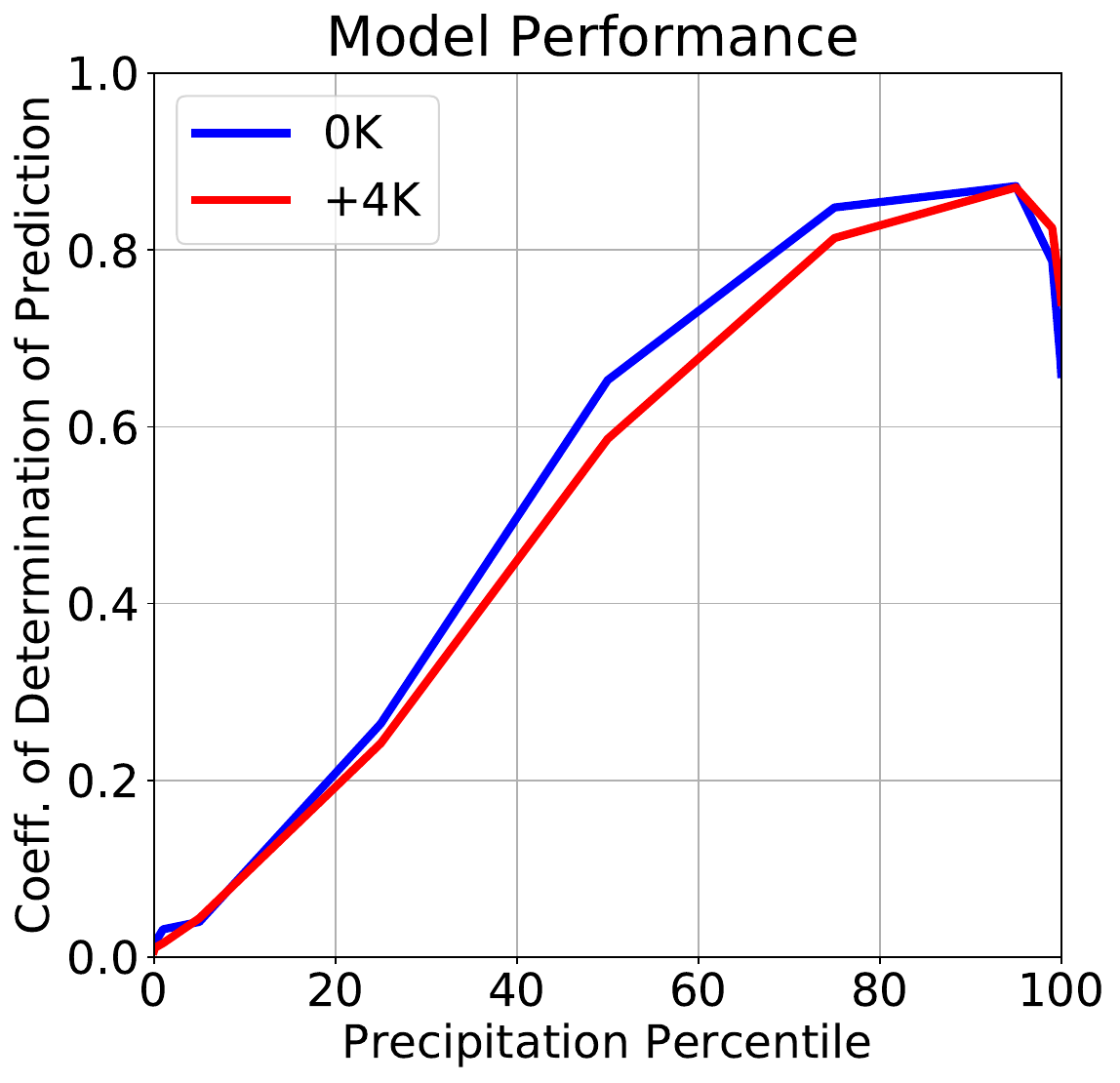}
\caption{\label{fig:model_scores} 
The simple results of the simple regression model we use to predict patterns in heavy precipitation ($\frac{P_{extrmeme}}{qsat}$) using just the dynamic contributions, $\pi_{Deep \ Convection}$ and $\pi_{Shallow \ Convection}$ identified by our unsupervised ML framework. We see our model works very well for high precipitation percentiles where the dynamic contributions are greatest and less well for lower percentiles where thermodynamics are also important.}
\end{figure}

\captionsetup[figure]{name={Figure S5}}
\begin{figure}[ht!]
\centering
\includegraphics[width=0.75\linewidth]{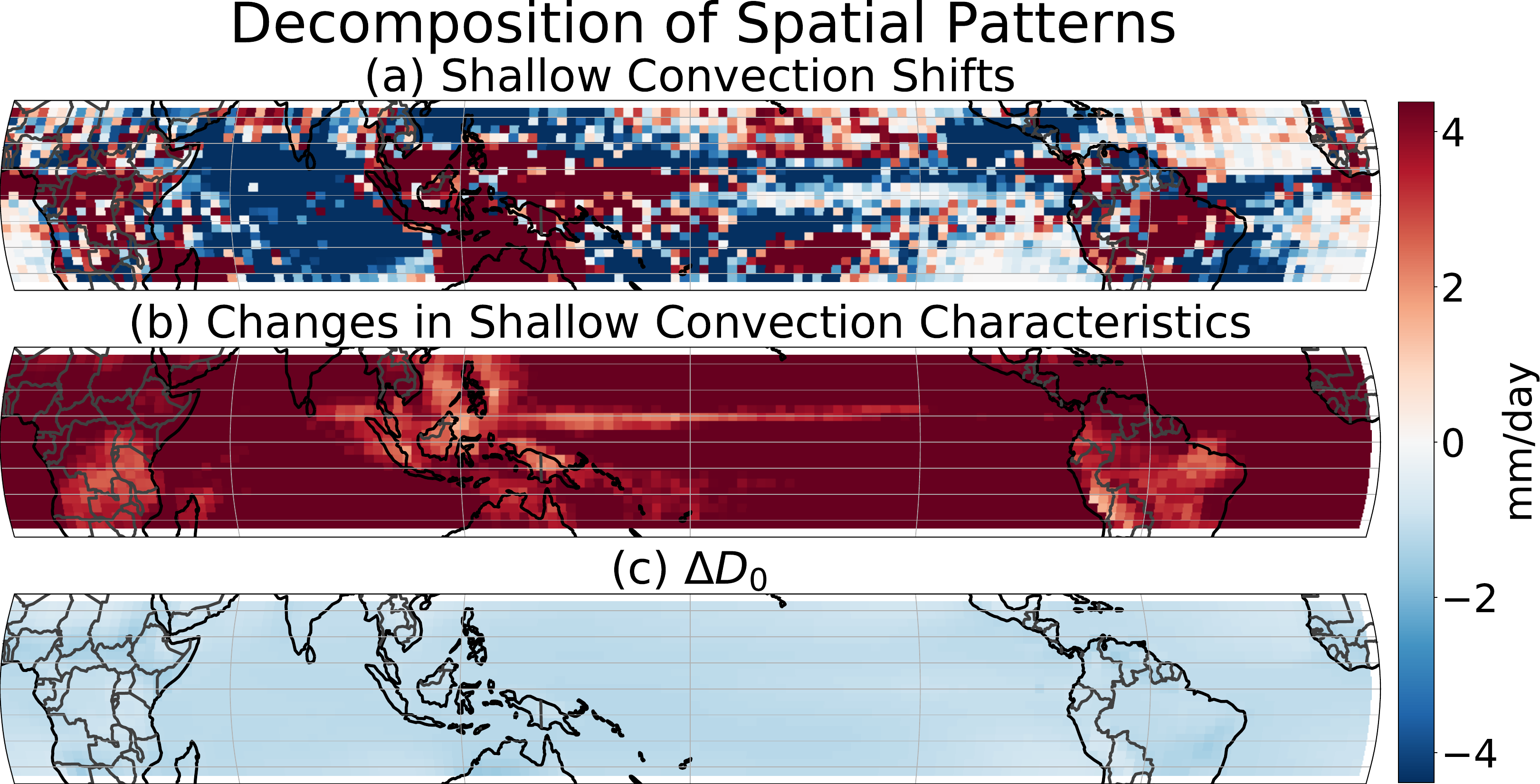}
\caption{\label{fig:SI_Decompostion} 
From Eq.~\ref{eq:dynamical_contribution}, we can decompose the changing spatial patterns (Fig~\ref{fig:Climate_Change}f) into five terms, including probability changes in shallow convection (a), changes in deep convective precipitation (b), and the intercept of Dynamical Prefactor (c).}
\end{figure}

\end{Backmatter}

\end{document}